\documentstyle[amsfonts,preprint,aps]{revtex}
\newtheorem{theorem}{Theorem}

\newcommand{\mybox}{\hfill{\scriptsize\fbox{\hspace*{1.1pt}}}}
\newcommand{\beq}{\begin{equation}}
\newcommand{\beqa}{\begin{eqnarray}} 
\newcommand{\eeqa}{\end{eqnarray}} 
\newcommand{\eeq}{\end{equation}}
\def\reals{\hbox{\rm I\kern -.2em R}}

\def\com{\mbox{$\mathbb{C}$}}
\def\dm{\mbox{$\cal{DM}$}}
\def\tr{\mbox{Tr}}

\def\dfn{\hbox{\rm $=$\kern -.95em $~^{^{\triangle}}$}}

\begin{document}

\title{GEOMETRY AND PRODUCT STATES}
\author{Robert B.Lockhart$^{1}$ \and Michael J. Steiner$^{2}$ \and Karl Gerlach$
^{2} $ \\
1 Mathematics Department, United States Naval Academy, Annapolis, Maryland
21401 \\
2 Naval Research Laboratory, Washington D.C.\\
}
\maketitle

\begin{abstract}
As separable states are a convex combination of product states, the geometry of the manifold of product states, $\Sigma$ is studied. Prior results by Sanpera, Vidal and Tarrach are extended. Furthermore, it is proven that states in the set tangent to $\Sigma$ at the maximally mixed state are separable; the set normal constains, among others, all maximally entangled states. A canonical decomposition is given. A surprising result is that for the case of two particles, the closest product state to the maximally entangled state is the maximally mixed state. An algorithm is provided to find the closest product state. 
\end{abstract}

\section{INTRODUCTION}

Entanglement has always been one of the quantum phenomena which sparked
debates about the completeness and interpretations of quantum mechanics \cite
{EPR},\cite{Bell},\cite{Omnes}. It is fundamental to
teleportation \cite{bennett}, \cite{werner}, secure key distribution, \cite
{shor},\cite{butler}, dense coding \cite{werner} and other applications.

Obviously, therefore, properties of entangled states, including ways to
determine if a state is entangled, are important. With regard to the latter,
several authors have given sufficient, necessary, and necessary and
sufficient conditions for a state to be entangled \cite{peres}, \cite
{horodecki}, \cite{chinese}, \cite{belgs}. With the exception of Peres's
partial transpose condition -- if any partial transpose of a state has a
negative eigenvalue, then the state is entangled -- these conditions tend to
be either impossible or hard to implement in all but low dimensional, low
rank cases. This, of course, is an important starting point for the
investigation of entanglement.

In this paper we take a different approach. Rather than trying to determine
if a state is entangled or not -- the latter case being known as separable
-- we study the whole set of separable states and the whole set of entangled
ones. Our approach is geometric and starts with the manifold of product
states, which we denote by $\Sigma $. One reason for starting with $\Sigma $
is that it is easy to decide if a density matrix is a product state by using
partial trace.  Another reason is that the totally mixed state $
\frac{1}{N}I$ is a product state. Lastly, $\Sigma $ is of interest in
itself. It is not convex or linear. But, as we shall show, it does have a
nice geometric property: if a straight line intersects $\Sigma $ in more
than two states, then every point on the line is a product state. Sanpera,
Vidal and Tarrach \cite{stv}proved this in the case of two qubits. We show
that it is true in general.

After establishing this geometric result, we turn to examining the set of
entangled and the set of separable states. Our approach is to start at the
totally mixed state, $\frac{1}{N}I$ and go in various directions. For
instance, we show that all states which are in the set tangent to $\Sigma $
at $\frac{1}{N}I$ are separable. On the other hand the maximally entangled
states are in the set which is normal to $\Sigma $ at $\frac{1}{N}I$.

One of our main results is the fact that in the case of $\com
^{n}\otimes \com^{n}$ the product state which is closest to a
maximally entangled state is $\frac{1}{N}I$, the totally mixed state. This
surprising result, which we shall, unfortunately, show is false for systems
with more than two particles, has an important consequence: there are no
product states inside the ball of radius $\sqrt{\frac{N-1}{N}}$ centered at
a maximally entangled state.

It may seem to the reader that it would be hard to determine if a state is
in the sets tangent or normal to $\Sigma $ at $\frac{1}{N}I$ and so
the results just mentioned may seem to be hard to use. However, this is not
the case. In fact it is easy to determine if a state is in the tangent or
normal set. One uses a canonical, orthogonal decomposition of $\tau
_{0}\left( N\right) $, the vector space of trace 0, Hermitian, $N\times N$
matrices, and the fact that every density matrix, $Q$, is uniquely
expressible as $Q=\frac{1}{N}I+H$, where $H\in \tau _{0}\left( N\right) $.

This fact and the canonical decomposition of $\tau _{0}\left( N\right) $ are
fairly standard mathematical fare; perhaps because $i\tau _{0}\left(
N\right) $ is the Lie Algebra $su(N)$. However, it might not be well-known
in the Quantum Information Community. Therefore, we shall present it in
detail for general multi-particle systems. To get an idea of what is
involved, consider the case of a density matrix $Q$ on $\com
^{n}\otimes \com^{m}$. It has the orthogonal decomposition $Q=\frac{1}{
N}I+H_{1}\otimes \frac{1}{m}I+\frac{1}{n}I\otimes H_{2}+H_{3}$, where $
H_{1}\in \tau _{0}\left( n\right) $, $H_{2}\in \tau _{0}\left( m\right) $,
and $H_{3}\in \tau _{0}\left( n\right) \otimes \tau _{0}\left( m\right) .$
It follows from results in \cite{lockhart}that $Q$ is in the tangent space
of $\Sigma $ at $\frac{1}{N}I$ if and only if $H_{3}=0$. Notice, it follows
from Theorem 3 in the present paper that if Q is entangled, then $H_{3}$ is
not zero. We see from this decomposition, therefore, conditions under which 
entanglement occurs. As for $Q$ being in the normal space of $\Sigma $ at $\frac{1}{
N}I$, that happens if and only if $H_{1}=H_{2}=0$. This follows, easily,
from the orthogonality of the decomposition. To actually find the
decomposition for a given state, one needs only to choose orthonormal bases
for $\tau _{0}\left( n\right) $ and $\tau _{0}\left( m\right) $. The set of
their tensor products then forms an orthonormal basis for $\tau _{0}\left(
n\right) \otimes \tau _{0}\left( m\right) $. Given these three orthonormal
bases, one then just takes inner products to get the decomposition.

We finish this paper with an algorithm which appears, in the bipartite case,
to give the product state which is closest to a given one.

\section{PRODUCT STATES}

We first establish some useful notation:

\begin{equation}
\tau _{k}\left( N\right) =\left\{ A\mid A\text{ is a Hermitian operator on }
\com^{N}\mbox{ and }  \tr(A)= k \right\},  \label{1.1}
\end{equation}
and
\begin{equation}
\dm(N)=\left\{ A\mid A\text{ is a density matrix on }\com^{N}\right\}  , \label{1.2}
\end{equation}
where $\tr$ denotes trace. Evidently, $\dm(N)$ is the subset of $\tau _{1}\left( N\right) $
consisting of positive, semi-definite operators. For $N=n_{1}\cdots n_{p},$
we need to describe the set of product state density matrices on $\com
^{N}=\com^{n_{1}}\otimes \cdots \otimes \com^{n_{p}}$. Using the
notation from \cite{lockhart}, we have the product operator

\begin{eqnarray}
\mu : \Pi _{1}^{p}\tau _{1}\left( n_{i}\right) & \rightarrow & \tau _{1}\left(
\Pi _{1}^{p}n_{i}\right) =\tau _{1}\left( N\right) \text{, \ \ \ given by}
\label{1.3} \\
\mu (A_{1},\ldots ,A_{p}) &=&A_{1}\otimes \cdots \otimes A_{p} . \nonumber 
\end{eqnarray}

The image of $\mu $ in $\tau _{1}(N)$, denoted by $\Sigma ,$ is a closed,
embedded submanifold \cite{lockhart}. In particular, every point in $\Sigma $
has a tangent space and normal space. We shall use this later, but first we
extend the result of Sanpera, Tarrach and Vidal\cite{stv} about lines.
Before doing so, we must point out that the set of product density matrices,
denoted $\Sigma ^{+}$, in $\dm\left( N\right) $ is just the image
of $\mu $ restricted to $\Pi_{1}^{p}{\cal DM}(n_{i})$ and so is a subset
of $\Sigma $. Thus the next theorem holds \textit{a fortiori }for product
density matrices.\ \ \ \ \ \ \ \ \ \ \ \ \ \ 

\begin{theorem}
If $A_{1}\otimes \cdots \otimes A_{p}=A$ and B $\in \tau _{1}\left( N\right) 
$, then one of the following is true about the line $r(t)=t\left(
A_{1}\otimes \cdots \otimes A_{p}\right) +\left( 1-t\right) B$:

\begin{enumerate}
\item  It intersects $\Sigma $ only at $A_{1}\otimes \cdots \otimes A_{p},$

\item  It intersects $\Sigma $ at $A_{1}\otimes \cdots \otimes A_{p}$ and
exactly one other point,

\item  The line lies in $\Sigma .$ In particular every density matrix on the
line is a product state.
\end{enumerate}
\end{theorem}

\noindent {\it Proof:} 

Suppose $B\neq A$ is also a product, thus $B=B_{1}\otimes \cdots \otimes
B_{p},$ and suppose $t_{0}A_{1}\otimes \cdots \otimes
A_{p}+(1-t_{0})B_{1}\otimes \cdots \otimes B_{p}=C_{1}\otimes \cdots \otimes
C_{p}$ for some $t_{0}\neq 0$ or $1$. In other words, suppose there are
three products on the line. We need to show $r(t)=tA_{1}\otimes
\cdots \otimes A_{p}+(1-t)B_{1}\otimes \cdots \otimes B_{p}$ is a product
for all t. If $A_{2}\otimes \cdots \otimes A_{p}=B_{2}\otimes \cdots \otimes
B_{p}$, then we have $r\left( t\right) =(tA_{1}+(1-t)B_{1})\otimes
A_{2}\otimes \cdots \otimes A_{p}$ and so are done. Therefore assume $
A_{2}\otimes \cdots \otimes A_{p}\neq B_{2}\otimes \cdots \otimes B_{p}$.
Note, since $\tr(A_{2}\otimes \cdots \otimes A_{p})=1=\tr(B_{2}\otimes
\cdots \otimes B_{p})$, we have that $A_{2}\otimes \cdots \otimes A_{p}\neq
B_{2}\otimes \cdots \otimes B_{p}$ is equivalent to them being linearly
independent.\ \ \ 

Let $\left\{ E_{i}\right\} $ be a basis for $\tau _{1}\left( n_{i}\right) $.
Then, using the Einstein summation convention of summing over repeated upper
and lower indices, we have 
\[
t_{0}A_{1}^{i}E_{i}\otimes A_{2}\otimes \cdots \otimes
A_{p}+(1-t_{0})B_{1}^{i}E_{i}\otimes B_{2}\otimes \cdots \otimes
B_{p}=C_{1}^{i}E_{i}\otimes C_{2}\otimes \cdots \otimes C_{p} .\]
Since the $E_{i}$ are linearly independent, this means that for each $i$ 
\[
t_{0}A_{1}^{i}A_{2}\otimes \cdots \otimes
A_{p}+(1-t_{0})B_{1}^{i}B_{2}\otimes \cdots \otimes
B_{p}=C_{1}^{i}C_{2}\otimes \cdots \otimes C_{p}. \]

If $C_{1}^{i_{0}}=0$ for some $i_{0}$, then $A_{1}^{i_{0}}=0=B_{1}^{i_{0}}$.
For otherwise we would have \ \ $t_{0}A_{1}^{i_{0}}A_{2}\otimes \cdots
\otimes A_{p}+(1-t_{0})B_{1}^{i_{0}}B_{2}\otimes \cdots \otimes B_{p}=0$,
which would contradict our assumption that $A_{2}\otimes \cdots \otimes
A_{p} $ and $B_{2}\otimes \cdots \otimes B_{p}$ are linearly independent.
Therefore assume $C_{1}^{i_{1}}\neq 0$ in which case we get 
\[
\frac{t_{0}A_{1}^{i_{1}}}{C_{1}^{i_{1}}}A_{2}\otimes \cdots \otimes A_{p}+
\frac{(1-t_{0})B_{1}^{i_{1}}}{C_{1}^{i_{1}}}B_{2}\otimes \cdots \otimes
B_{p}=C_{2}\otimes \cdots \otimes C_{p} . \]

It follows from this and the linear independence of $A_{2}\otimes \cdots
\otimes A_{p}$ and $B_{2}\otimes \cdots \otimes B_{p}$ that $\frac{
t_{0}A_{1}^{i_{1}}}{C_{1}^{i_{1}}}=\frac{t_{0}A_{1}^{i_{2}}}{C_{1}^{i_{2}}}$
and $\frac{(1-t_{0})B_{1}^{i_{1}}}{C_{1}^{i_{1}}}=\frac{
(1-t_{0})B_{1}^{i_{2}}}{C_{1}^{i_{2}}}$ for every $i_{1}$, $i_{2}$ such that 
$C_{1}^{i_{1}}$ and $C_{1}^{i_{2}}$ are not $0$.

Since we already know that $A_{1}^{i}=B_{1}^{i}=0$ if $C_{1}^{i}=0$, we may
conclude for all $i$ that $A_{1}^{i}=\frac{A_{1}^{i_{1}}}{C_{1}^{i_{1}}}
C_{1}^{i}$ and $B_{1}^{i}=\frac{B_{1}^{i_{1}}}{C_{1}^{i_{1}}}C_{1}^{i}$,
where $C_{1}^{i_{1}}\neq 0$. Thus $A_{1}=\frac{A_{1}^{i_{1}}}{C_{1}^{i_{1}}}
C_{1}=\lambda C_{1}$ and $B_{1}=\frac{B_{1}^{i_{1}}}{C_{1}^{i_{1}}}C_{1}=\mu
C_{1}$. Since $\tr(A_{1})=\tr(B_{1})=\tr(C_{1})=1$, we have $\lambda
=\mu =1$ and so $A_{1}=B_{1}=C_{1}$. This means we may write 
\[
t_{0}A_{1}\otimes \cdots \otimes A_{p}+(1-t_{0})B_{1}\otimes \cdots \otimes
B_{p}=C_{1}\otimes \cdots \otimes C_{p} \] as \[
A_{1}\otimes (t_{0}A_{2}\otimes \cdots \otimes A_{p}+(1-t)B_{2}\otimes
\cdots \otimes B_{p})=A_{1}\otimes C_{2}\otimes \cdots \otimes C_{p}  .
\] By induction we have that $r(t)=tA_{1}\otimes \cdots \otimes
A_{p}+(1-t)B_{1}\otimes \cdots \otimes B_{p}$ is a product for all t. Hence
if there are three matrices on the line $r(t)$ which are products,
then all are products.

This actually proves the theorem, though one wants to know if the other two
cases (one product or two products) can occur. In the next theorem we show
that there are lines with only one product matrix on them. As for there
being lines with only two, if that did not occur then every two elements in $
\Sigma $ would be connected by a line in $\Sigma $. This would mean $\Sigma $
is convex. In particular every separable state would be a product state.
However, this cannot be, for it was shown in \cite{lockhart} that $\Sigma
^{+}$ has measure $0$ in $\dm(N)$; but the set of separable states
has an open interior and so is not measure 0. Thus there are lines with
exactly two product states on them
\mybox
% \end{proof}

\bigskip

Before proceeding, we need to say a few words about the set of vectors
tangent to $\Sigma $. Since $\Sigma $ is the image of the embedding $\mu $,
it follows that the tangent space at $A=A_{1}\otimes \cdots \otimes A_{p}$
is $d\mu (T_{(A_{1},...A_{p})})$, where $T_{(A_{1},...A_{p})}$ is the
tangent space of $\Pi _{1}^{p}\tau _{1}(n_{i})$ at $(A_{1},...A_{p})$ and $
d\mu $ is the Jacobian of $\mu $. Thus the tangent space of $\Sigma $ at $
A=A_{1}\otimes \cdots \otimes A_{p}$ is the vector space spanned by

\begin{equation}
\label{eqn4}
\left\{ 
\begin{array}{c}
H_{1}\otimes A_{2}\otimes \cdots \otimes A_{p},A_{1}\otimes H_{2}\otimes
A_{3}\otimes \cdots \otimes A_{p}, \\ 
...,A_{1}\otimes \cdots \otimes A_{p-1}\otimes H_{p}:H_{i}\in \tau
_{0}\left( n_{i}\right)
\end{array}
\right\} .
\end{equation}

Note, we distinguish between the tangent space of $\Sigma $ at $A$ and the
set tangent to $\Sigma $ at $A.$ The former is the vector space of
vectors which are tangent to $\Sigma $ at $\frac{1}{N}I$ and so provide the
directions which are tangent to $\Sigma $ at $A.$ The latter is the
set  of states which are obtained by starting at $A$ and going in a
tangential direction. Thus if $\mathcal{T}_{A}$ denotes the tangent space of 
$\Sigma $ at $A$, then the set tangent to $\Sigma $ at $A$ is $A+
\mathcal{T}_{A}$. Similarly, the normal space of $\Sigma $ at $A$ is the
vector space of vectors normal to $\Sigma $ at $A$ and the set normal
to $\Sigma $ at $A$ is the set  of states which are obtained by
starting at $A$ and going in a normal direction. If $\mathcal{N}_{A}$
denotes the normal space, then the set normal to $\Sigma $ at $\ A$
is $A+\mathcal{N}_{A}$.

\begin{theorem}
Let $A=A_{1}\otimes \cdots \otimes A_{p}$ be such that each $A_{i}=\frac{1}{
n_{i}}I$ for $i\neq i_{0}$. If the line $r(t)=tR+(1-t)A$ is
orthogonal to the tangent space of $\Sigma $ at $A$, then $A$ is the only
product matrix on the line. \ \ \ \ \ \ \ \ \ 
\end{theorem}

\noindent {\it Proof:}

Suppose $Q=Q_{1}\otimes \cdots \otimes Q_{p}$ is also on the line. Since the
line and tangent space are orthogonal, we have for all $j$ and all $H_{j}\in
\tau _{0}\left( n_{j}\right) $

\begin{eqnarray}
0 &=&\left\langle Q-A,A_{1}\otimes \cdots \otimes H_{j}\otimes \cdots
A_{p}\right\rangle =  \label{1.5.1} \\
&&\left\langle Q_{1},A_{1}\right\rangle \cdots \left\langle
Q_{j},H_{j}\right\rangle \cdots \left\langle Q_{p},A_{p}\right\rangle
-\left\langle A_{1},A_{1}\right\rangle \cdots \left\langle
A_{j},H_{j}\right\rangle \cdots \left\langle A_{p},A_{p}\right\rangle . \nonumber
\end{eqnarray}

Now note that for $i\neq i_{0}$ we have $\left\langle
Q_{i},A_{i}\right\rangle =\left\langle Q_{i},\frac{1}{n_{i}}I\right\rangle =
\frac{1}{n_{i}}\tr Q_{i}=\frac{1}{n_{i}},$ and $\left\langle
A_{i},A_{i}\right\rangle =\left\langle \frac{1}{n_{i}}I,\frac{1}{n_{i}}
I\right\rangle =\frac{1}{n_{i}}$. Thus when $\ j=i_{0}$, (\ref{1.5.1}) becomes $
\left\langle Q_{i_{0}},H_{i_{0}}\right\rangle =\left\langle
A_{i_{0}},H_{i_{0}}\right\rangle $ for all $H\in \tau _{0}\left(
n_{i_{0}}\right) $. This can only happen if $Q_{i_{0}}=A_{i_{0}}$. However,
if $j\neq i_{0}$ then (\ref{1.5.1}) reduces to $\left\langle
Q_{i_{0}},A_{i_{0}}\right\rangle \left\langle Q_{j},H_{j}\right\rangle =0$
for all $H_{j}\in \tau _{0}\left( n_{j}\right) $, since in this case $A_{j}=
\frac{1}{n_{j}}I$ and $\left\langle \frac{1}{n_{j}}I,H_{j}\right\rangle =0$.
Thus $\left\langle Q_{j},H_{j}\right\rangle =0$ for all $H_{j}$ and so $
Q_{j}=\frac{1}{n_{j}}I=A_{j}$ for all $j\neq i_{0}.$ \mybox
% \end{proof}

We finish this section by showing every density matrix in the set 
tangent to $\Sigma $ at the totally mixed state is separable.

\begin{theorem}
Suppose $Q$ is a density matrix and $Q$ is in the set tangent to $
\Sigma $ at $\frac{1}{N}I$, then $Q$ is separable.
\end{theorem}

\noindent {\it Proof:}

For $Q$ to be in the set tangent to$\ \Sigma $ at $\frac{1}{N}I$, $Q$
must be expressible as 

\begin{eqnarray}
Q &=&\frac{1}{N}I+H_{1}\otimes \frac{1}{n_{2}}I\otimes \cdots \otimes \frac{1
}{n_{p}}I+\frac{1}{n_{1}}I\otimes H_{2}\otimes \cdots \otimes \frac{1}{n_{p}}
I+\cdots \\
&&+\frac{1}{n_{1}}I\otimes \cdots \otimes \frac{1}{n_{p-1}}I\otimes H_{p} . \nonumber
\end{eqnarray}

If $\mid \psi _{i}\rangle $ is an eigenvector of $H_{i}$ with eigenvalue $
\lambda _{i}$, then it is easy to see $\mid \psi _{1}\cdots \psi _{p}\rangle 
$ is an eigenvector of $Q$ with eigenvalue $\frac{1}{N}(1+\sum_{1}^{p}n_{i}
\lambda _{i})$, where, as above, $N=n_{1}\cdots n_{p}$. Since each $H_{i}$
has trace equal to 0, the minimum eigenvalue of each $H_{i}$ is 0, in which
case $H_{i}=0$, or is negative. Let $\mu _{i}$ be the minimum eigenvalue of $
H_{i}$ for each $i$. Then $\mu =\frac{1}{N}(1+\sum_{1}^{p}n_{i}\mu _{i})$ is
the minimum eigenvalue of $Q$. Since $Q$ is a density matrix it is positive
semi-definite and so $0\leq \mu $. This in turn implies $\frac{1}{N}I+\frac{1
}{n_{1}}I\otimes \cdots \otimes H_{i}\otimes \cdots \otimes \frac{1}{n_{p}}I$
is positive semi-definite and so a density matrix. In fact, if $\mu _{i}\neq
0$, then $\frac{1}{N}I+\frac{1}{n_{1}}I\otimes \cdots \otimes F_{i}\otimes
\cdots \otimes \frac{1}{n_{p}}I$, where $F_{i}=\frac{1}{n_{i}\left| \mu
_{i}\right| }H_{i}$, is a density matrix, since the most negative eigenvalue
of $F_{i}$ is $\frac{\mu _{i}}{n_{i}\left| \mu _{i}\right| }=-\frac{1}{n_{i}}
$.

Noting that $\frac{1}{N}I+\frac{1}{n_{1}}I\otimes \cdots \otimes
F_{i}\otimes \cdots \otimes \frac{1}{n_{p}}I=\frac{1}{n_{1}}I\otimes \cdots
\otimes (\frac{1}{n_{i}}I+F_{i})\otimes \cdots \otimes \frac{1}{n_{p}}I$, we
see that $Q$ is separable, for it is the convex combination 
\[
(1-\sum_{1}^{p}n_{i}\left| \mu _{i}\right| )\frac{1}{N}I+\sum_{1}^{p}n_{i}
\left| \mu _{i}\right| (\frac{1}{n_{1}}I\otimes \cdots \otimes (\frac{1}{
n_{i}}I+F_{i})\otimes \cdots \otimes \frac{1}{n_{p}}I) . \] \mybox

% \end{proof}

\section{A CANONICAL, ORTHOGONAL DECOMPOSITION FOR DENSITY MATRICES}

In this section we present a very useful orthogonal decomposition of trace
1, Hermitian matrices in terms of the totally mixed state and trace 0,
Hermitian matrices. To begin, note that if $Q$ is a trace 1, Hermitian
matrix on $\com^{N}$ then $Q-\frac{1}{N}I$ $=S$ has trace 0 and so is
in the vector space $\tau _{0}\left( N\right) .$ Also note that a trace 0
matrix is orthogonal to $\frac{1}{N}I$, since $\left\langle \frac{1}{N}
I,S\right\rangle =\frac{1}{N}Tr(S)=0$. Thus we need to decompose $\tau
_{0}\left( N\right) $, where $N=n_{1}\cdots n_{p},$ into orthogonal
subspaces.

Let $\mathbb{Z}\mbox{$_{2}= \left\{ 0,1\right\}.$}$ The product of p-copies of $
\mathbb{Z}\mbox{$_{2}$}$ is denoted $\mathbb{Z}\mbox{$_{2}^{p}$}$. It is the set of all
strings of length p of zeros and ones. For $\alpha =\left( \alpha
_{1},\ldots \alpha _{p}\right) \in \mathbb{Z}\mbox{$_{2}^{p}$}$ take $\left| \alpha
\right| =\sum_{1}^{p}\alpha _{i}.$ Thus $\left| \alpha \right| $ is the
number of ones in $\alpha .$ For $\alpha \in \mathbb{Z}\mbox{$_{2}^{p}$},$ take $\nu
(\alpha )$ to be the vector space $\nu (\alpha )=\nu _{1}(\alpha )\otimes
\cdots \otimes \nu_{p} \left( \alpha \right) ,$ where $\nu _{j}(\alpha )=\frac{1
}{n_{j}}I$ if $\alpha _{j}=0$ and $\nu _{j}(\alpha )=\tau _{0}(n_{j})$ if $
\alpha _{j}=1.$ With these notational conventions we have,

\[
\tau _{0}\left( N\right) =\oplus _{k=1}^{p}\oplus _{\left| \alpha \right|
=k,\alpha \in \mathbb{Z}\mbox{$_{2}^{p}$}}\nu (\alpha ) . \] That this is an orthogonal decomposition of $\tau _{0}\left( N\right) $
follows from the fact that if $A=A_{1}\otimes \cdots \otimes A_{p}$ and $
B=B_{1}\otimes \cdots \otimes B_{p}$, then $\left\langle A,B\right\rangle
=\prod_{i=1}^{p}\left\langle A_{i},B_{i}\right\rangle $ and the fact that $
\left\langle \frac{1}{N}I,H\right\rangle =0$ if $H$ is trace 0.

As can be seen from (\ref{eqn4}) the subspace with $k=1$, forms the
directions which are tangential to $\Sigma $ at $\frac{1}{N}I$. Thus $Q$ is
in the set normal to $\Sigma $ at $\frac{1}{N}I$ if $Q=\frac{1}{N}I+S$
and $S\in \oplus _{2}^{k}\oplus _{\left| \alpha \right| =k}\nu (\alpha )$.

To close this section we mention that the only role trace 1 plays in this
decomposition is to determine the scalar multiple of $I$. In particular, any
observable can be similarly decomposed.

\section{STATES THAT HAVE THE TOTALLY MIXED ONE AS THE BEST PRODUCT STATE
APPROXIMATION}

Suppose $Q\in \tau _{1}\left( N\right) $, where $N=n_{1}\cdots n_{p}$. If $P$
is the closest point on $\Sigma $ to $Q$, then the line joining $Q$ to $P$
is perpendicular to the tangent space of $\Sigma $ at $P.$ Thus the only
density matrices which could have the totally mixed state as best product
state approximation (i.e. have $\frac{1}{N}I$ as the closest product state)
lie in the set normal to $\Sigma $ at $\frac{1}{N}I$. We showed in
the last section how to characterize this set and shortly we
shall show the maximally entangled states lie in it. But first we extend
Theorem 3 by using the decomposition in the last section.

Set ${\cal T}=\oplus_{\left| \alpha \right| =1}\nu (\alpha )$ and $
{\cal N}=\oplus_{k=2}^{p}\oplus _{\left| \alpha \right| =k}\nu (\alpha )$
.

\begin{theorem}
Every density matrix, $Q,$ on $\com^{n_{1}}\otimes \cdots \otimes 
\com^{n_{p}}$ is uniquely expressible as $Q=\frac{1}{N}I+Q_{\mathcal{T}
}+Q_{\mathcal{N}},$ where $Q_{\mathcal{T}}\in \mathcal{T}$ and $Q_{\mathcal{N
}}\in \mathcal{N}$ . In fact this is an orthogonal decomposition of $Q$. If $
Q$ is entangled, then $Q_{\mathcal{N}}\neq 0.$ If the totally mixed state is
the closest product state to $Q$, then $Q_{\mathcal{T}}=0.$ If $Q_{\mathcal{N
}}=0$, \ then $Q$ is in the set tangent to $\Sigma $ at $\frac{1}{N}I$
. If $Q_{\mathcal{T}}=0$, then $Q$ is in the set normal to $\Sigma $
at $\frac{1}{N}I$.
\end{theorem}

Suppose $N=n^{p}$ and $\mid \psi _{i}\rangle $ is an orthonormal basis for $
\com^{n}$. Modulo, $U(n)\otimes \cdots \otimes U(n)$ action (i.e.
local operations), the maximally entangled state associated with this
orthonormal basis is the projection onto $\frac{1}{\sqrt{n}}\sum_{1}^{n}\mid
\psi _{i}...\psi _{\iota }\rangle ,$ with $\mid \psi _{i}...\psi _{i}\rangle 
$ being the p-fold tensor product of $\mid \psi _{i}\rangle $. Using
standard tensor analysis notation, this projection is 
\begin{equation}
E_{j_{1}\cdots j_{p}}^{i_{1}\cdots i_{p}}=\left\{ 
\begin{array}{l}
1/n\text{ }\ \ if\text{\ }i_{1}=\cdots =i_{p}\text{ and }j_{1}=\cdots =j_{p},
\\ 
0\text{ \ \ \ \ \ otherwise. }
\end{array}
\right.
\end{equation}

For what follows we need to compute $\left\langle E,B\right\rangle $ for an
arbitrary trace 1, Hermitian matrix, $B$. Using the decomposition in the
previous section, we know that $B$ \ is the orthogonal sum of matrices $
B(\alpha ),$ $\alpha \in \mathbb{Z}\mbox{$_{2}^{p}$}$, where $B(\alpha )=B_{1}(\alpha
)\otimes \cdots \otimes B_{p}(\alpha )$, with $B_{j}\left( \alpha \right) =
\frac{1}{n}I$ if $\alpha _{j}=0$ and $B_{j}(\alpha )\in \tau _{0}\left(
n\right) $ if $\alpha _{j}=1$. Thus we only need to compute $\left\langle
E,B(\alpha )\right\rangle $, since $\left\langle E,B\right\rangle $ is the
sum of such quantities.

Since $B\left( \alpha \right) $ is a product $(B(\alpha ))_{t_{i}\ldots
t_{p}}^{s_{1}\ldots s_{p}}=(B_{1}(\alpha ))_{t_{1}}^{s_{1}}\cdots
(B_{p}(\alpha ))_{t_{p}}^{s_{p}}$. Again using the Einstein summation
convention and using the fact $E_{j_{1}\cdots j_{p}}^{i_{1}\cdots i_{p}}=0$
unless $i_{1}=\cdots i_{p}$ and $j_{1}=\cdots j_{p},$ we obtain

\begin{eqnarray}
\label{eqn8}
\left\langle E,B\left( \alpha \right) \right\rangle &=& E_{j_{1}\cdots
j_{p}}^{i_{1}\cdots i_{p}}(B_{1}(\alpha ))_{i_{1}}^{j_{1}}\cdots
(B_{p}(\alpha ))_{i_{p}}^{j_{p}} , \\
&=&E_{j\cdots j}^{i\cdots i}(B_{1}(\alpha ))_{i}^{j}\cdots (B_{p}(\alpha
))_{i}^{j}  . \nonumber
\end{eqnarray}

When $\left| \alpha \right| <p$ at least one of the $\alpha _{j}=0$ and so
at least one of the $B_{j}\left( \alpha \right) =\frac{1}{n}I$. For this one
we have $(B_{l}(\alpha ))_{i}^{j}=\left\{ 
\begin{array}{l}
0\text{ if }i\neq j \\ 
\frac{1}{n}\text{ if }i=j
\end{array}
\right. $. Thus in this case (\ref{eqn8}) becomes

\begin{eqnarray}
\label{eqn9}
\left\langle E,B\left( \alpha \right) \right\rangle &=&E_{i\cdots
i}^{i\cdots i}(B_{1}(\alpha ))_{i}^{i}\cdots (B_{p}(\alpha ))_{i}^{i} , \\
&=&\frac{1}{n}(B_{1}(\alpha ))_{i}^{i}\cdots (B_{p}(\alpha ))_{i}^{i} . \nonumber
\end{eqnarray}

\begin{theorem}
The maximally entangled state $E$ is in the set normal to $\Sigma $ at $
\frac{1}{N}I.$
\end{theorem}

\noindent {\it Proof:} 

What it means for $E$ to be in the set normal to $\Sigma $ at $\frac{1
}{N}I$ is that the vector parallel to the line which connects $\frac{1}{N}I$
to $E$ is perpendicular to any vector tangent to $\Sigma $ at $\frac{1}{N}I.$
We have seen above that these tangent vectors are of the form $\
\sum_{\left| \alpha \right| =1}B\left( \alpha \right) .$ Hence we need to
show that $\left\langle E-\frac{1}{N}I,\sum_{\left| \alpha \right|
=1}B\left( \alpha \right) \right\rangle =0.$ This follows because

\begin{equation}
\label{eqn10}
\left\langle E-\frac{1}{N}I,\sum_{\left| \alpha \right| =1}B\left( \alpha
\right) \right\rangle =\sum_{\left| \alpha \right| =1}(\left\langle
E,B\left( \alpha \right) \right\rangle -\frac{1}{N}\left\langle I,B\left(
\alpha \right) \right\rangle
\end{equation}
and all the terms on the right hand side of (\ref{eqn10}) are zero. Indeed the term $
\frac{1}{N}\left\langle I,B\left( \alpha \right) \right\rangle =0,$ since $
B\left( \alpha \right) $ is trace 0. And the $\left\langle E,B\left( \alpha
\right) \right\rangle $ are 0 because $\left| \alpha \right| =1$. When $
\left| \alpha \right| =1,$ there is only one factor in $B\left( \alpha
\right) $ which is not $\frac{1}{n}I$, and so $\left\langle E,B\left( \alpha
\right) \right\rangle $\ reduces to $\frac{1}{n^{p}}(B_{j}\left( \alpha
\right) )_{i}^{i}$, with $B_{j}\left( \alpha \right) $ being the one factor
which is not $\frac{1}{n}I$. But $(B_{j}\left( \alpha \right) )_{i}^{i}$ is
the trace of $B_{j}\left( \alpha \right) $, which is 0. \mybox
% \end{proof}

Because of this theorem, we know there is the possibility that the closest
product state to $E$ is $\frac{1}{N}I$. We shall now show that is in fact
the case for two particles, but not the case for more than two. First of all
we note that the distance squared from $E$ to any other state $C$ is

\bigskip 
\begin{equation}
\label{eqn11}
\left\langle E-C,E-C\right\rangle =\left\| E\right\| ^{2}-2\mbox{Re}
\left\langle E,C\right\rangle +\left\| C\right\| ^{2} .
\end{equation}

When $C$ is the totally mixed state, $\frac{1}{N}I$, this reduces to $
\left\| E\right\| ^{2}-\frac{1}{N}.$ We need to compare this with the
distance from $E$ to a product state. Suppose $A$ is a product state. Then $
A=(\frac{1}{n}I+R_{1})\otimes \cdots \otimes \left( \frac{1}{n}
I+R_{p}\right) $,where $R_{j}\in \tau _{0}\left( n\right) $. Expanding this
expression for $A$, we get $A=\frac{1}{N}I+\sum_{k=1}^{p}\sum_{\left| \alpha
\right| =k,}B(\alpha )$, where for each $\alpha \in \mathbb{Z}\mbox{$_{2}^{p}$}$ we
have $B(\alpha )=B_{1}(\alpha )\otimes \cdots \otimes B_{p}(\alpha ),$ with $
B_{j}(\alpha )=\frac{1}{n}I$ if $\alpha _{j}=0$ and $B_{j}(\alpha )=R_{j}$
if $\alpha _{j}=1.$ \ Recall from above that the $B(\alpha )$ are mutually
orthogonal and there are \small $\left( \begin{array}{c} p \\ k \end{array} \right)$ \normalsize of them for each $k$. Set $
B_{k}=\sum_{\left| \alpha \right| =k}B(\alpha )$. The $B_{k}$ are also
mutually orthogonal. In terms of them, $A=\frac{1}{N}I+\sum_{k=1}^{p}B_{k}$.

Using the mutual orthogonality of the $B_{k}$, we find

\begin{eqnarray*}
\left\langle E,A\right\rangle &=&\left\langle E,\frac{1}{N}I\right\rangle
+\sum_{k=1}^{p}\left\langle E,B_{k}\right\rangle , \\
&=&\frac{1}{N}+\sum_{k=1}^{p}\left\langle E,B_{k}\right\rangle .
\end{eqnarray*}
Similarly, $\left\| A\right\| ^{2}=\frac{1}{N}+\sum_{k=1}^{p}\left\|
B_{k}\right\| ^{2}$. Hence it follows from (\ref{eqn11}) that the distance
squared from $E$ to $A$ is

\begin{equation}
\label{eqn12}
\left\langle E-A,E-A\right\rangle =\left\| E\right\| ^{2}-\frac{1}{N}-2\mbox{Re}\sum_{k=1}^{p}\left\langle E,B_{k}\right\rangle +\sum_{k=1}^{p}\left\|
B_{k}\right\| ^{2} .
\end{equation}

Recalling that the distance squared from $E$ to the totally mixed state is $
\left\| E\right\| ^{2}-\frac{1}{N}$, we see that the totally mixed state is
the closest product state to $E$ if and only if $\sum_{k=1}^{p}(\left\|
B_{k}\right\| ^{2}-2\mbox{Re}\left\langle E,B_{k}\right\rangle )\geq 0$ for
all choices of $R_{1},...,R_{p}\in \tau _{0}\left( n\right) $ such that $A$
is a density matrix. Let us first compute $\left\langle E,B_{k}\right\rangle 
$. For $k<p$ we can use (\ref{eqn9}) to get

\begin{eqnarray}
\label{eqn13}
\left\langle E,B_{k}\right\rangle &=&\sum_{\left| \alpha \right|
=k}\left\langle E,B\left( \alpha \right) \right\rangle , \\
&=&\sum_{\left| \alpha \right| =k}\frac{1}{n}(B_{1}(\alpha ))_{i}^{i}\cdots
(B_{p}(\alpha ))_{i}^{i}  . \nonumber
\end{eqnarray}

Suppose for $\alpha =(\alpha _{1},...,\alpha _{p})$ that $\alpha _{j}=1$ for 
$j=m_{1},...,m_{k}$. Then the summand in (\ref{eqn13}) becomes $\frac{1}{n^{p-k+1}}
\left( R_{m_{1}}\right) _{i}^{i}\cdots (R_{m_{k}})_{i}^{i}$. Thus for $k<p$
we have

\begin{equation}
\left\langle E,B_{k}\right\rangle =\frac{1}{n^{p+1-k}}\sum_{m_{1}<\cdots
<m_{k}}(R_{m_{1}})_{i}^{i}\cdots (R_{m_{k}})_{i}^{i} .
\end{equation}

For $k=p$ we have $\left\langle E,B_{p}\right\rangle
=E_{j...j}^{i...i}(R_{1})_{i}^{j}\cdots (R_{p})_{i}^{j}=\frac{1}{n}
\sum_{i,j}(R_{1})_{i}^{j}\cdots (R_{p})_{i}^{j}$. Hence

\begin{eqnarray}
\label{eqn14}
2\mbox{Re}\sum_{k=1}^{p}\left\langle E,B_{k}\right\rangle &=&\frac{2}{n^{p+1}
}\sum_{k=1}^{p}\sum_{m_{1}<\cdots <m_{k}}(nR_{m_{1}})_{i}^{i}\cdots
(nR_{m_{k}})_{i}^{i} \\
&&+\frac{2}{n}\mbox{Re}\sum_{i\neq j}(R_{1})_{i}^{j}\cdots (R_{p})_{i}^{j} .
\nonumber 
\end{eqnarray}
Dispensing with the Einstein summation notation from here on out, we can
rewrite (\ref{eqn14}) as

\begin{eqnarray}
\left\langle E,B_{k}\right\rangle &=& \\
&&\frac{2}{n^{p+1}}(\sum_{i=1}^{n}(\prod_{l=1}^{p}(1+n(R_{l})_{i}^{i}))-1))+
\frac{2}{n}\mbox{Re}\sum_{i\neq j}(R_{1})_{i}^{j}\cdots (R_{p})_{i}^{j} . \nonumber
\end{eqnarray}

We also need to compute $\left\| B_{k}\right\| ^{2}$. Fortunately, this is
quite simple since the $B(\alpha )$ are orthogonal. In particular, we have
for all $k$ that

\begin{eqnarray}
\left\| B_{k}\right\| ^{2} &=&\sum_{\left| \alpha \right| =k}\left\|
B(\alpha )\right\| ^{2}=\sum_{\left| \alpha \right| =k}\left\| B_{1}(\alpha
)\right\| ^{2}\cdots \left\| B_{p}(\alpha )\right\| ^{2} , \\
&=& \frac{1}{n^{p-k}}\sum_{m_{1}<\cdots <m_{k}}\left\| R_{m_{1}}\right\|
^{2}\cdots \left\| R_{m_{k}}\right\| ^{2} . \nonumber
\end{eqnarray}

Thus

\begin{eqnarray}
\sum_{k=1}^{p}\left\| B_{k}\right\| ^{2} &=&\frac{1}{n^{p}}
\sum_{k=1}^{p}\sum_{m_{1}<\cdots <m_{k}}n\left\| R_{m_{1}}\right\|
^{2}\cdots n\left\| R_{m_{k}}\right\| ^{2} , \\
&=&\frac{1}{n^{p}}(\prod_{l=1}^{p}(1+n\left\| R_{l}\right\| ^{2})-1) . \nonumber
\end{eqnarray}

We can now rephrase our question as follows: is there a choice of $R_{j}\in
\tau _{0}(n)$ such that $A=(\frac{1}{n}I+R_{1})\cdots (\frac{1}{n}I+R_{p})$
is a density matrix and

\bigskip 
\begin{eqnarray}
\label{eqn19}
&&\frac{1}{n^{p}}(\prod_{l=1}^{p}(1+n\left\| R_{l}\right\| ^{2})-1) \\
&&-\frac{2}{n^{p+1}}(\sum_{i=1}^{n}(\prod_{l=1}^{p}(1+n(R_{l})_{i}^{i}))-1))-
\frac{2}{n}\mbox{Re}\sum_{i\neq j}(R_{1})_{i}^{j}\cdots (R_{p})_{i}^{j} \nonumber
\end{eqnarray}
is negative. If so, then that product state is closer to the maximally
entangled state than the totally mixed one. If not, then the totally mixed
state is the closest product state to the maximally entangled one.

\begin{theorem}
For a quantum system modelled on $\com^{n}\otimes \com^{n}$ the
closest product state to a maximally entangled state is the totally mixed
one. Because of this there are no product states within $\sqrt{\frac{N-1}{N}}
$, where $N=n^{2},$ of a maximally entangled state. For quantum systems with
more than two particles, the totally mixed state is not the closest product
state to a maximally entangled state.
\end{theorem}

\noindent {\it Proof: }

Let us first consider the case of two particles. Thus $p=2$ in (\ref{eqn19}) and so
it reduces to

\begin{eqnarray}
\label{eqn20}
&&\frac{1}{n^{2}}(1+n\left\| R_{1}\right\| ^{2})(1+n\left\| R_{2}\right\|
^{2})-\frac{1}{n^{2}} \\
&&-\frac{2}{n^{3}}\sum_{i=1}^{n}((1+n(R_{1})_{i}^{i})(1+n(R_{2})_{i}^{i})-1)
\nonumber \\
&&-\frac{2}{n}\mbox{Re}\sum_{i\neq j}(R_{1})_{i}^{j}(R_{2})_{i}^{j} . \nonumber
\end{eqnarray}

Expanding the first two terms, cancelling, regrouping and using the fact the 
$R_{l}$ are Hermitian and so $(R_{2})_{i}^{j}=\overline{\left( R_{2}\right) }
_{j}^{i}$ we see that (\ref{eqn20}) equals

\begin{eqnarray}
&&\frac{1}{n}\left( \left\| R_{1}\right\| ^{2}+\left\| R_{2}\right\|
^{2}-2[(R_{1})_{i}^{i}(R_{2})_{i}^{i}+\mbox{Re}\sum_{i\neq j}(R_{1})_{i}^{j}
\overline{(R_{2})}_{j}^{i}]\right) \\
&&+\left\| R_{1}\right\| ^{2}\left\| R_{2}\right\| ^{2}-\frac{2}{n^{2}}
((R_{1})_{i}^{i}+(R_{2})_{i}^{i}) . \nonumber
\end{eqnarray}

Since the $R_{l}$ are trace 0, the last term in this expression is 0. On the
other hand the first term is $\frac{1}{n}\left\langle R_{1}-\overline{R_{2}}
,R_{1}-\overline{R_{2}}\right\rangle $, which is greater than or equal to
zero, with equality only if $R_{1}=R_{2}=0.$ Since the same is true for the
middle term in (\ref{eqn20}), we have proved that $\frac{1}{N}I$ is the product state
closest to $E$ in the bipartite case.

Now suppose $p\geq 3$ and take all the $R_{l}$ to be the matrix $R$ with diagonal elements:

\[
R_{i}^{j}=\left\{ 
\begin{array}{l}
1/2n\text{ \ \ \ \ if }i=j=1 \\ 
-1/2n\text{ }\ \ \text{if }i=j=2 \\ 
0\text{ \ \ \ \ \ \ \ \ \ otherwise.}
\end{array}
\right. \]

When (\ref{eqn19}) has this $R$ substituted into it, it becomes

\[
\frac{1}{n^{p}}\left( \left( 1+\frac{1}{2n}\right) ^{p}+\frac{4}{n}-1-\frac{2
}{n}((\frac{3}{2})^{p}+(\frac{1}{2})^{p})\right) . \]

For $p=3$, this reduces to $8n^{3}(-12n^{2}+6n+1),$ which is negative for $
n\geq \dot{2}.$ Since for such $n$ the quantity ($1+\frac{1}{2n})$ is less
than $\frac{3}{2},$ for fixed $n\geq 2$, the expression $\left( 1+\frac{1}{2n
}\right) ^{p}+\frac{4}{n}-1-\frac{2}{n}((\frac{3}{2})^{p}+(\frac{3}{2})^{p})$
decreases with increasing $p$ and so is negative for all $p\geq 3.$

In conclusion, the geometry of states was studied. Prior results by Sanpera, Vidal and Tarrach were extended and several new results were presented.  These results are useful in the characterization of entangled states, which has been a difficult problem due to the complicated structure of the entangled space.  
\vspace{.1in}

\noindent{{\bf Acknowledgement:}} The authors acknowledge the Office of Naval Research for the support of this work.

\section{APPENDIX}

In this Appendix an algorithm is presented to find the closest product state to an arbitary state. It was used, for example, to initially find that the closest product state to the maximally entangled state is the maximally mixed state. The algorithm converges very rapidly (often within a few iterations) for modest $N$, where $N=n^{2}$. 

Given a density matrix $C\in \com^{n}\otimes\com^{n}$, consider the problem of finding 
\[ \mbox{arg~} \min_{A,B} || C-A\otimes B|| \]
where $A,B\in \com^{n}$ are density matrices. An outline of the algorithm is first provided, and later the details on the computation of $A_{o}$ and $B_{o}$ are given.  

\subsection{Algorithm}

\begin{enumerate}
\item Set $B= I_{n}$, . 
\item Given $B$, find $A_{o}$ as computed in Subsection B.
\begin{itemize}
\item Set $A= A_{o}+ \frac{1}{n}(1-\tr(A_{o})) I_{n}$
\item Check if $A$ is positive semi-definite (psd). If yes, then $A$ satisfies a Lagrange Multiplier equation (see (\ref{eqnlagr}) and discussion) skip to Step 3. If no, an $A$ that satisfies a Lagrange Multiplier equation (see (\ref{eqnsix}) and discussion) can be found as follows:
\begin{itemize}
\item Compute the eigenvalues of $A_{o}$, represented by  $\lambda_{1}\ge\lambda_{2}\cdots \ge \lambda_{n}$ and let $\Phi_{1},\Phi_{2},\cdots,\Phi_{n}$ be corresponding eigenvectors. 
\item Set $M=\mbox{arg} \min_{m} \{ m|(n-1)\lambda_{j} + \lambda_{m} + 1 - \tr(A_{o}) \ge 0, j=1,2,\cdots,n; j\ne m \}$. 
\item For $m=1,2,\cdots,M$, compute $A_{m} = A_{o} + \frac{1}{n-1}(1-\tr(A_{o})+\lambda_{m}) I_{n} - \frac{1}{n-1}(1-\tr(A_{o})+n \lambda_{m})\Phi_{m} \Phi_{m}'$
\item Set  $A= \mbox{arg} \min_{A_{m}} ||C-A_{m}\otimes B||_{2} $
\end{itemize}
\end{itemize}
\item Given $A$ find $B_{o}$. This is accomplished by considering the element transform $\mbox{T}$ such that $\mbox{T}(A \otimes B)$ = $B \otimes A$.
Setting $\tilde{C} = \mbox{T}(C)$, it is seen that $|| C-A\otimes B|| =|| \tilde{C}-B\otimes A||$. Then apply the methodology as in Steps 1. and 2. above with the appropriate changes to the transformed problem. 
\item Check $||C-A\otimes B||$ for convergence. If no, go to Step 2.
\end{enumerate}

\subsection{Solution for $A_{o}$ and $B_{o}$}
\label{subs}

Finding $A_{o}$ is equivalent to finding $a_{ij}$ $(i,j=1,2,\cdots,n)$ such that 
\[ a_{ij}= \mbox{arg} \min_{a_{ij}} || C_{ij}- a_{ij} B|| . \]
Now
\[ || C_{ij}-a_{ij} B||^{2} = \tr \{ (C_{ij}-a_{ij}B)(C_{ij}-a_{ij}B)'\} \]
where $'$ denotes conjugate transpose.  The above is minimized when
\begin{eqnarray}
\label{eqnaa}
a_{ij}^{o} & = & \frac{\tr(C_{ij} B')}{||B||^{2}} \mbox{~~~} i\ne j , \\ \nonumber
a_{ii}^{o} & = & \frac{1}{2} \frac{\tr (C_{ii}B')+ \tr(BC_{ii}')}{||B||^{2}} .
\end{eqnarray}
Finding $B_{o}$ can be easily accomplished with the same result as above, by considering the element transform $\mbox{T}$ such that $\mbox{T}(A \otimes B)$ = $B \otimes A$.
Setting $\tilde{C} = \mbox{T}(C)$, it is seen that
\[ || C-A\otimes B|| =|| \tilde{C}-B\otimes A||,\] and the solution for $b_{ij}^{o}$ is by symmetry with (\ref{eqnaa}) given by
\begin{eqnarray*}
b_{ij}^{o} & = & \frac{\tr(\tilde{C}_{ij} A')}{||A||^{2}} \mbox{~~~} i\ne j , \\
b_{ii}^{o} & = & \frac{1}{2} \frac{\tr (\tilde{C}_{ii}A')+ \tr(A\tilde{C}_{ii}')}{||A||^{2}} .
\end{eqnarray*}

\noindent {\bf Remark} {\it The solution to $A$ without the constraint $\tr(A)=1$ for \[ \mbox{arg~} \min_{A} || C-A\otimes B|| \] where $B,C$ are psd Hermitian (psdh) matrices, can be shown to be psdh. Hence both $A_{o}$ and $B_{o}$ are always psd. }

Consider now the solution to $A$ with $\tr(A)=1$ for \[ \mbox{arg~} \min_{A} || C-A\otimes B|| \] where $B,C$ are given density matrices. The Lagrange multiplier solution to \beq \label{eqnlagr} \bigtriangledown_{A} \{ ||C-A\otimes B||_{2} + \gamma \tr(A) \} =0 \eeq 
can be shown to have the form $A=A_{o} + \tilde{\gamma} I_{n} $, where $\bigtriangledown_{A}$ is the gradient, $\bigtriangledown_{A} \dfn \frac{\partial}{\partial A}$. Substituting this form in the constraint equation, $\tr(A)=1$ gives $\tilde{\gamma}=1-\tr(A_{o})$.  Since this is a convex minimization problem, an interior point solution to Lagrange multiplier gradient equations is guaranteed to be the minimum. If $A$ is psd ($A$ is Hermitian), then we will use this solution in the algorithm, although strictly speaking, it should be checked to see if it is an interior point. 

The solution to $A$ with $\tr(A)=1$ for \[ \mbox{arg~} \min_{A\in \mbox{psdh}} || C-A\otimes B|| \] where $B,C$ are density matrices is not as straightforward if $A=A_{o} + (1-\tr(A_{o})) I_{n}$ is not psd.  In this case, we use the additional constraints $x'Ax \ge 0$ and $x'x=1$. The minimization problem is no longer convex. However, if the minimum is an interior point, then it necessarily satisfies the Lagrange multiplier gradient equations. As $A=A_{o} + (1-\tr(A_{o})) I_{n}$  is not psd but still minimizes $||C-A\otimes B||_{2}$ , then the solution lies on the boundary of the constraint $x'Ax\ge 0$, i.e. $A$ is singular, hence $x'Ax=0$.
Now, the Lagrange multiplier problem becomes the solution to:
\beq \label{eqngrad1} \bigtriangledown_{A} \{ ||C-A\otimes B||_{2} + \gamma_{1}\tr(A) + \gamma_{2} x'A x + \gamma_{3} x'x \} = 0\eeq
\beq \label{eqngrad2} \bigtriangledown_{x} \{ ||C-A\otimes B||_{2} + \gamma_{1}\tr(A) + \gamma_{2} x'A x + \gamma_{3} x'x \} = 0. \eeq
Eqn (\ref{eqngrad2}) above merely shows that $x$ is an eigenvector of $A$. Equation (\ref{eqngrad1}) imples that $A$ has the following form:
\[ A= A_{o} + \tilde{\gamma_{1}}I_{n} + \tilde{\gamma_{2}}x x' .\]
Now, $\tilde{\gamma_{1}},\tilde{\gamma_{2}}$ must satisfy constraint equations:
\beq  \tr(A) = 1 = \tr(A_{o}) + n \tilde{\gamma_{1}} + \tilde{\gamma_{2}} \eeq
and  \begin{eqnarray} \label{eqnthr}   0  & = &  A x \nonumber \\ 
 & = & A_{o} x + \tilde{\gamma_{1}} x + \tilde{\gamma_{2}} x \nonumber \\
\label{eqntwo} & = & (A_{o} + (\tilde{\gamma_{1}}  + \tilde{\gamma_{2}})I_{n}) x .  
\end{eqnarray}
Equation (\ref{eqnthr}) implies that $x'$ is an eigenvector of $A_{o}$. Denote the eigenvalues of $A_{o}$ by $\lambda_{1}\ge\lambda_{2}\cdots \ge \lambda_{n}$ and $\Phi_{1},\Phi_{2},\cdots,\Phi_{n}$ the corresponding eigenvectors.  Let $x=\Phi_{m}$, then $Ax=0$ implies 
\beq
\label{eqnfour} 
\tilde{\gamma_{1}} + \tilde{\gamma_{2}} = - \lambda_{m}(A_{o}). \eeq
Using (\ref{eqntwo}) and (\ref{eqnfour}) we can show
\beq \label{eqnfive} A = A_{o} + \frac{1}{n-1}(1-\tr(A_{o})+\lambda_{m}) I_{n} - \frac{1}{n-1}(1-\tr(A_{o})+n \lambda_{m})\Phi_{m} \Phi_{m}'. \eeq
We can show using (\ref{eqnfive}) that $x'A x= \Phi_{m}' A \Phi_{m} = 0$. It would appear, we have $n$ possible candidates for the solution of $A$: each solution would be based on using a different $\Phi_{m}$ in (\ref{eqnfive}). However, we still must ensure that $A$ is psd. The solutions for $A$ are Hermitian.

In order for $A$ to be psd, we required that the eigenvalues of $A$ (denoted by $\gamma_{j}, j=1,2,\cdots, n$) to satisfy $\gamma_{j}\ge 0$. Now, we can show if $x= \Phi_{m}$ then using (\ref{eqnfive}) the following $n-1$ conditions must be satisfied:
\beq \label{eqnsix} (n-1)\gamma_{j} = (n-1) \lambda_{j} + \lambda_{m} + 1-\tr(A_{o}) \ge 0, \mbox{~for~} j=1,2,\cdots, n, j\ne m .\eeq
Note $\lambda_{1}= \max_{j} \lambda_{j}$. If the minimum occurs at an interior point then we are guaranteed that a solution for $A$ must exist and \[ (n-1)\lambda_{j} + \lambda_{1} + 1-\tr(A_{o}) \ge (j-1) \lambda_{j} + \lambda_{m} + 1 - \tr(A_{o}) \] for all $j,m$. Therefore \[ (n-1)\lambda_{j} + \lambda_{1} + 1 - \tr(A_{o}) \ge 0 \] for all $j$.
Thus we will successively check the largest eigenvalues of $A_{o}$ and see which satisfy (\ref{eqnsix}) (again $\lambda_{1}$ must satisfy (\ref{eqnsix}) if an interior point minimum exists). The solution for $A_{o}$ can be checked numerically to see if it is an interior point minimum.

\bigskip

\bigskip

\end{document}